\documentclass[twocolumn,english,prl,amssymb,aps,superscriptaddress,showpacs,twocolumn,amsmath,showkeys,floatfix]{revtex4-1}
\usepackage[T1]{fontenc}
\usepackage[latin9]{inputenc}
\usepackage{geometry}
\usepackage[active]{srcltx}
\usepackage{textcomp}
\usepackage{amsmath}
\usepackage{graphicx}
\usepackage{color}
\usepackage{esint}
\usepackage{svg}
\usepackage{siunitx}
\usepackage{physics}
\usepackage[version=4]{mhchem}
\makeatletter
\usepackage{babel}

\makeatother

\begin{document}
	
	\title {Non-symmetrical vortex beam shaping in VECSEL laser arrays}
	\author{Sopfy Karuseichyk}
	\affiliation{Universit\'e Paris-Saclay, CNRS, ENS Paris-Saclay, CentraleSup\'elec, LuMIn, 91190 Gif-sur-Yvette, France}
	\author{Ilan Audoin}
	\affiliation{Universit\'e Paris-Saclay, CNRS, ENS Paris-Saclay, CentraleSup\'elec, LuMIn, 91190 Gif-sur-Yvette, France}
	\author{Vishwa Pal}
	\affiliation{Department of Physics, Indian Institute of Technology Ropar, Rupnagar, Punjab 140001, India}
	\author{Fabien Bretenaker}
	\affiliation{Universit\'e Paris-Saclay, CNRS, ENS Paris-Saclay, CentraleSup\'elec, LuMIn, 91190 Gif-sur-Yvette, France}

	\begin{abstract} 
We propose and numerically test a novel concept for asymmetric vortex beam generation in a Degenerate Vertical External Cavity Surface Emitting Laser (DVECSEL). The method is based on a phase-locking ring array of lasers created inside a degenerate cavity with a binary amplitude mask containing circular holes. The diffraction engineering of the mask profile allows to control the complex coupling between the lasers. The asymmetry between different lasers is introduced by varying the hole diameters corresponding to different lasers. Several examples of masks with non-uniform or uniform circular holes are investigated numerically and analytically to assess the impact of non-uniform complex coupling coefficients on the degeneracy between the vortex and anti-vortex steady-states of the ring laser arrays. It is found that the in-phase solution always dominates irrespective of non-uniform masks. The only solution to make one particular vortex solution dominant over other possible steady-state solutions consists in imprinting the necessary phase shift among neighboring lasers in the argument of their coupling coefficients. We also investigate the role of the Henry factor inherent to the use of a semiconductor active medium in the probabilities to generate vortex solutions. Analytical calculations are performed to generalize a  formula previously reported in Opt. Express {\bfseries 30}, 15648 (2022) for the limiting Henry factor to cover the case of complex couplings. 	\end{abstract}
	
	%\pacs{??}
	
	\maketitle
	
\section{Introduction}
 A particular category of structured light beams, called optical vortices (OV), have been discovered a few decades ago, and since then are considered for many applications. Such OV beams exhibit a phase singularity at the center, around which the phase accumulated by the field is an integer multiple of $2\pi$. %The phase of the OV corresponds to a discrete number of the $2\pi$ phase accumulation around the phase-singularity point in the center of the beam. 
This integer number is known as the topological charge (TC)/orbital angular momentum (OAM) of OV beams. The annular intensity distribution of OV has been found useful for optical trapping \cite{itsatrap, Vpartcl, Vtwez}, micromotoring \cite{Vspanner, massTranspotrV}, microscopy \cite{STED}, nano-structuring \cite{PhysRevLett.110.143603}, and optical data transfer \cite {Wang2012, Gibson:04,OAMcommun,Vencript}. 

In many of these applications, it was shown that improved performance with additional control could be achieved by introducing intensity and/or phase asymmetry in the beam, which then becomes an asymmetric optical vortex (AOV) beam. For example, the rate of microparticle motion was shown to increase linearly with the asymmetry of vortex-carrying Bessel- or Laguerre-Gaussian beams \cite{10.1063/1.4958309, Kovalev:16}. The thermal damage of the live cells was also reported to be better managed when the symmetry of OV is broken \cite{10.1063/1.4958309}.

For some particular applications, this maximization of the vortex beam asymmetry may be detrimental to some extent parameters. For example, an experiment involving particle trapping would become less efficient in terms of particle conservation if the involved OV is asymmetric. A trade-off must also be found when one considers OAM-based optical data transfer \cite {Wang2012, Gibson:04}. The vortex-based information storage using the topological charge can be improved using some additional information encoded in the non-symmetrical intensity profile \cite{BeamCod}. % Indeed, the shape of the beam is conserved through propagation and thus constitutes a unique parameter, similar to the phase profile of the beam. {\color{red}[NOT clear??]}. 
A large asymmetry can allow the beam to increase its storage capacity. However, a strong deformation of the OV can make the information carried by the phase profile more difficult to read. The asymmetric aberration laser beams have also been demonstrated for obtaining additional control on generating high-energy density at desired spatial locations \cite{Singh:2022}. Consequently, the flexibility and adaptive control of the beam profile are critical for all these applications. 

The most common solutions for AOV beam generation are the optical system misalignment, the use of digital micromirror device (DMD), spatial light modulator (SLM), or a pinhole plate located on the laser beam's path \cite{10.1063/1.5024445, DMD, Kotlyar:14JOSA, Hsieh:18, Fries2018-el,Li:13}. In most of these works, the AOV beams are generated with a single laser and thus can pose power limitations for various applications. Further, the generation of AOV beams external to the laser source using DMD and SLM can further reduce the output power due to filtration of higher diffraction orders containing considerable power. To overcome these limitations as well as to get high-purity beams, the most promising approach is adaptive beam-shaping integrated directly into the laser source \cite{Forbes:2019}. To this aim, the dynamics of OV generation in degenerate cavity solid-state lasers has been broadly studied. In particular, the self-healing properties of OVs generated by phase-locked laser arrays as well as probability of generating different topological charges have been thoroughly investigated \cite{Pal2015, Pal2017, Dev2021,Piccardo2022, Forbes:2019}.

The probability to generate OV with certain TC was predicted to be also quite large in degenerate cavity lasers based on semiconductor gain chips, namely VECSELs (Vertical External Cavity Surface Emitting Laser). Moreover, this probability was shown to depend on the value of the Henry-factor in such gain structures \cite{first}. Additionally, these lasers are known for their smooth class-A synchronization dynamics \cite{first} and very low noise operation \cite{De:13}. The gain chip production technology allows to choose the best resonant wavelength in the near-infrared range for any particular application. Thus degenerate VECSELs are now considered to be promising solutions for all of the above-mentioned applications.

The aim of the present paper is thus to theoretically explore the dynamics of non-symmetrical OV generation using a ring array of lasers formed in a degenerate VECSEL. The rotational symmetry of the system is broken by altering the geometry of the amplitude-mask, which forms a ring array of lasers. Different kinds of asymmetric masks are investigated, namely point defects, gradient in diameter, and random hole diameters. The difference of such masks with respect to uniform hole parameters is expected to affect the coupling coefficient between the lasers and thus affect the dynamics of the system and the choice of the system steady-state. Moreover, the influence of the Henry factor, which is particular to semiconductor gain media \cite{first}, on the probability to achieve phase-locking of the laser array with non-symmetrical OV phase differences is studied in details.  

\section{Description of a laser array with asymmetric coupling}
\begin{figure}[ht]
\centering
   \includegraphics[width=0.8\linewidth]{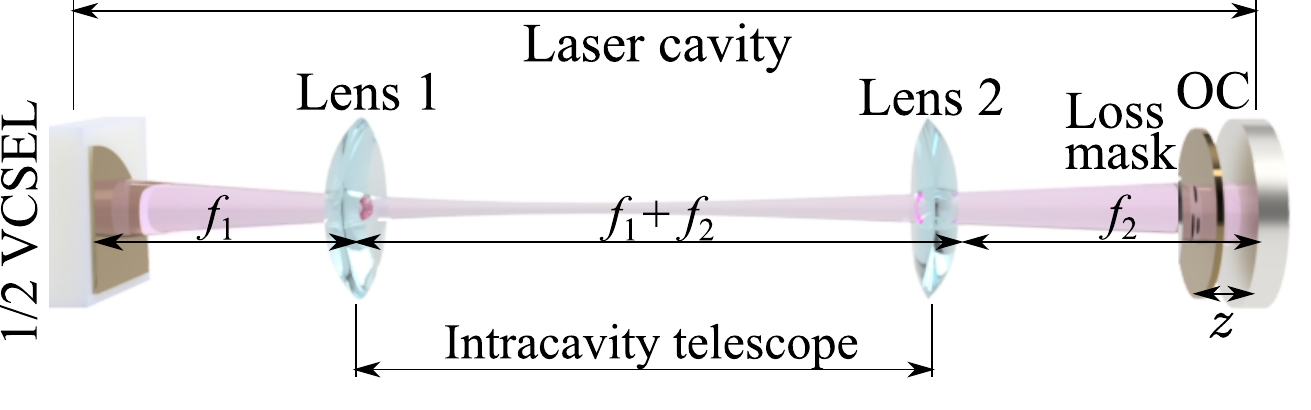}
    \caption{Scheme of a DVECSEL. The 1/2-VCSEL gain chip and amplitude mask are placed at the near-field planes of the intracavity telescope. The telescope is formed by two lenses ($L_1$ and $L_2$ with focal lengths $f_{1}$ and $f_{2}$, respectively) in self-imaging positions. OC represents the output coupler. %\color{red} [ADD PUMP SOURCE TO THE SCHEMATIC]
    }
    \label{fig:sh}
\end{figure}

We investigate AOV generation using a ring array of lasers based on a degenerate cavity VECSEL. The scheme of such a laser system is shown in Fig.\,\ref{fig:sh}. The semiconductor gain chip typically consists of a Bragg reflector and a few optically pumped quantum wells, thus acting as an amplifying planar mirror. The cavity is closed by a second planar mirror, and its self-imaging properties are obtained by inserting two lenses forming a telescope. An amplitude mask is inserted close to the output coupler (OC), i.~e. close to the self-imaging point of the intracavity telescope. In this configuration, every hole in the mask defines an individual laser. The gap between the mask and output coupler has a length of $z$ and influences the amount of laser light injected into neighboring apertures due to diffraction at the mask's edges.

For the modeling, we rely on the values of the parameters of our recent experiments \cite{Karuseichyk:23} in which $f_1 = 5\,\mathrm{cm}$, $f_2 = 20\,\mathrm{cm}$ and the holes in the mask form a regular circular pattern with diameter of $200\,\si{\um}$ and center-to-center separation between them of $250\,\si{\um}$.

\begin{figure}[ht]
    \centering  
    \includegraphics[width=0.8\linewidth]{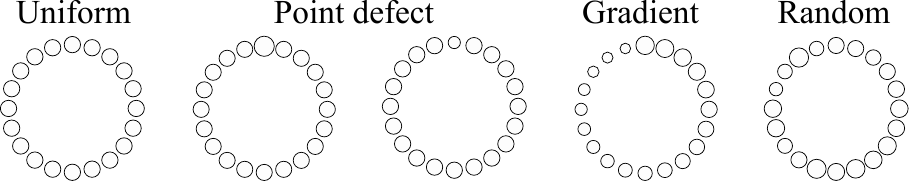}
    \caption{Schematic shapes of the different amplitude-mask patterns, called Uniform,  Point defect, Gradient, and Random. The hole diameters are not to scale.}
    \label{fig:masks}
\end{figure}

A mask pattern in which the holes have identical diameters will be called a ``Uniform" mask geometry. In the present paper, we investigate the generation of AOV beams by introducing different kinds of asymmetries in the array, called ``Gradient", ``Random", and ``Point defect" masks, as presented in Fig.\,\ref{fig:masks}. 
In the Point defect case, one of the holes has a smaller or bigger diameter than the others. In the Gradient case, the hole diameters increase gradually from the first to the last one. Finally, in the random configuration, the hole diameters are randomly chosen around some average value.

The hole diameters in Fig.\,\ref{fig:masks} are not to scale with the array geometry. In agreement with experiments \cite{first}, the standard hole diameter is taken to be 200\,$\,\si{\um}$ and the center-to-center separation between two successive holes is $d=250\,\,\si{\um}$. %Variation of the diameters falls within the range $0.25 - 3\,\mu m$. 
In the following, we consider arrays of 20 lasers. The details of the different masks are given in Table \ref{tab:data}. %

\begin{table}[ht]
    \centering
  \caption{\bf Hole diameters in different chosen configurations of 20 lasers$^\textit{a}$}    \begin{tabular}{c c c c c c }
       \hline 
       Hole index & 1 & 2 & 10 & 19 & 20   \\
           \hline      
        ``Big" defect  & 202 & 200 & 200 & 200 & 200\\
        ``Small" defect & 197 & 200 & 200 & 200 & 200 \\
        Gradient & 197 & 197.25 & 199.25 & 201.5 & 201.75 \\
        Random & 197.25 & 198.25 & 201 & 199.5 & 200\\
   \hline
   \end{tabular}
    \label{tab:data}
    
    $^\textit{a}$ The standard deviation of the hole diameters in the random configuration is chosen equal to $1.5\,\si{\um}$.
\end{table}

%Random mask - {197.25, 198.25, 201.75, 198.5, 198., 199., 197.75, 201.25, 199.75, 201., 197.5, 199.25, 200.75, 201.5, 200.25, 198.75, 200.5, 197., 199.5, 200.}

\section{Coupling between neighboring lasers}
\begin{figure}[ht]
\centering
    \includegraphics[width=0.8\linewidth]{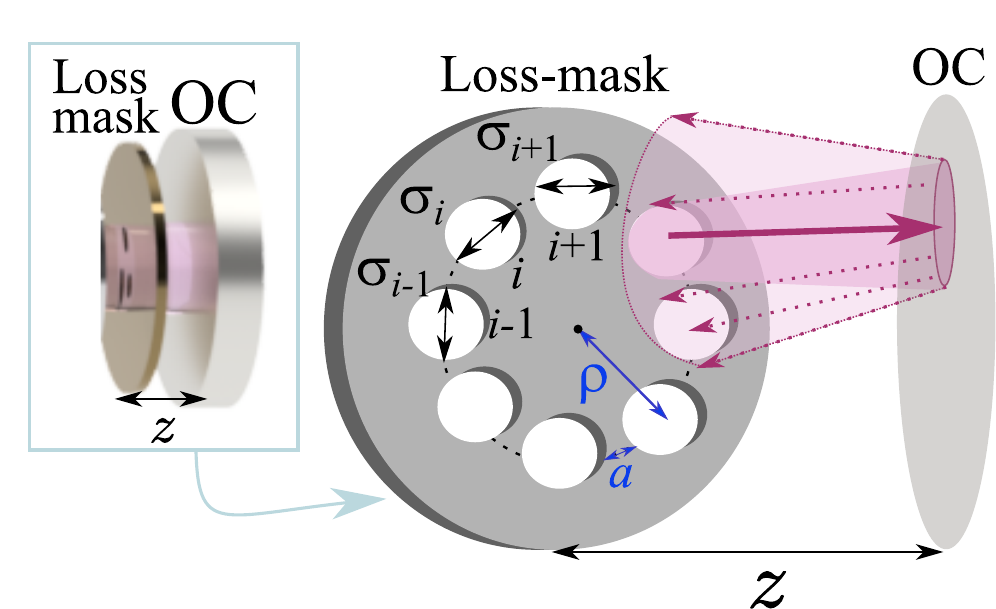}\\
    \caption{Definition of the mask parameters. This
    example corresponds to $n = 8$ holes. The pink area represents the fraction of the beam coming from one hole injected into its neighbors after reflection on the output coupler (OC). %caption and diffraction
    }
    \label{fig:masks_2}
\end{figure}

At every round-trip inside the cavity, each laser undergoes diffraction by its mask hole, and after reflection on the planar mirror, a fraction of the diffracted light is injected into its neighbors. This process is schematically shown in Fig.\,\ref{fig:masks_2}. 
The corresponding coupling parameter is obtained by calculating the overlap between the diffracted laser field and the neighboring laser mode profiles. To calculate the wavefront diffracted by the laser labeled by $i$, we suppose that the incident field on the circular hole has a uniform amplitude $E_{0,i}$. Then, after propagating along the distance $2\,z$, i.~e., twice the distance between the mask and the output coupler, the diffracted beam profile is obtained by the following Huygens-Fresnel equation in cylindrical coordinates as
\begin{align}
        E_{s,i}(\rho') =\frac{\text{i}\pi E_{0,i}}{ z\lambda}\int \limits_{0}^{\sigma_{i}/2}\rho%\exp{\displaystyle\frac{-\text{i}\pi (\rho^2+\rho'^2)}{2 z\lambda}}
        \exp{\frac{-\text{i}\pi (\rho^2+\rho'^2)}{2 z\lambda}}\nonumber\\\times 
        J_0\left(\frac{\pi\rho \rho'}{ z\lambda}\right)\textrm{d}\rho\ ,\label{eq:1}
\end{align}
% z - is one way. The formula is already simplified in terms of 2z - for the round-trip.
where $E_{s,i}$ is the field diffracted at a distance $2z$ from the $i^{\textrm{th}}$ hole with diameter $\sigma_{i}$, and where $\rho$ and $\rho'$ are the radial coordinates in the planes of the incident and diffracted fields, respectively. The VECSEL wavelength is taken to be $\lambda=1\,\si{\um}$. 

The diffracted waveform exhibits strong oscillations, whose characteristics mainly depend on the hole diameter $\sigma_i$ and on the propagation distance $z$. The fraction of the field from laser number $i$ injected by diffraction to laser number $i\pm1$ at each round trip is a complex coupling coefficient, whose value can be calculated by projecting the field diffracted by hole $i$ on laser $i\pm1$ as

\onecolumngrid\
\begin{equation}
        \kappa_{i\rightarrow i\pm1} =\frac{ \displaystyle \int\limits_0^{2\pi}\textrm{d}\phi_{i\pm1}\int\limits_0^{\sigma_{i\pm1}/2}\rho_{i\pm1} \textrm{d}\rho_{i\pm1}E_{0,i\pm1}^*(\rho_{i\pm1})E_{s,i}\left(\sqrt{d^2+\rho_{i\pm1}^2+2d\rho_{i\pm1}\cos\phi_{i\pm1}}\right)}{\sqrt{\displaystyle \int\limits_0^{2\pi}\textrm{d}\phi_i\int\limits_0^{\sigma_i/2}\rho_i \textrm{d}\rho_i |E_{0,i}(\rho_i)|^2\int\limits_0^{2\pi}\textrm{d}\phi_{i\pm1}\int\limits_0^{\sigma_{i\pm1}/2}\rho_{i\pm1} \textrm{d}\rho_{i\pm1} |E_{0,{i\pm1}}(\rho_{i\pm1})|^2}}\ .
        \label{eq:kapa}        
\end{equation}
\noindent\rule{\linewidth}{0.4pt}
\vspace{0.1em}
\twocolumngrid\
In Eq.\,(\ref{eq:kapa}), $(\rho_{i},\phi_{i})$ and $(\rho_{i\pm 1},\phi_{i\pm 1})$ are the cylindrical coordinates centered on apertures number $i$ and $i\pm 1$, respectively. We have also supposed that all the laser modes have a cylindrical symmetry and thus their fields depend only on their radial coordinates $\rho_i$ and $\rho_{i\pm 1}$. 
We treat the incident mode profiles on the mask holes as top hat profiles with a width equal to $\sigma_i$ and in a range given by Table\,\ref{tab:data}. Thus, the denominator of Eq.\,(\ref{eq:kapa}) simply becomes $\pi\sigma_i\sigma_{i\pm 1}|E_{0,i}E_{0,i\pm 1}|/4$. The calculations of the overlap with more realistic field profiles would add some scaling factor to it's modulus value, but would not affect the coupling argument. 

\begin{figure}
\centering
   \vspace{1.6em}
   \includegraphics[width=0.8\linewidth]{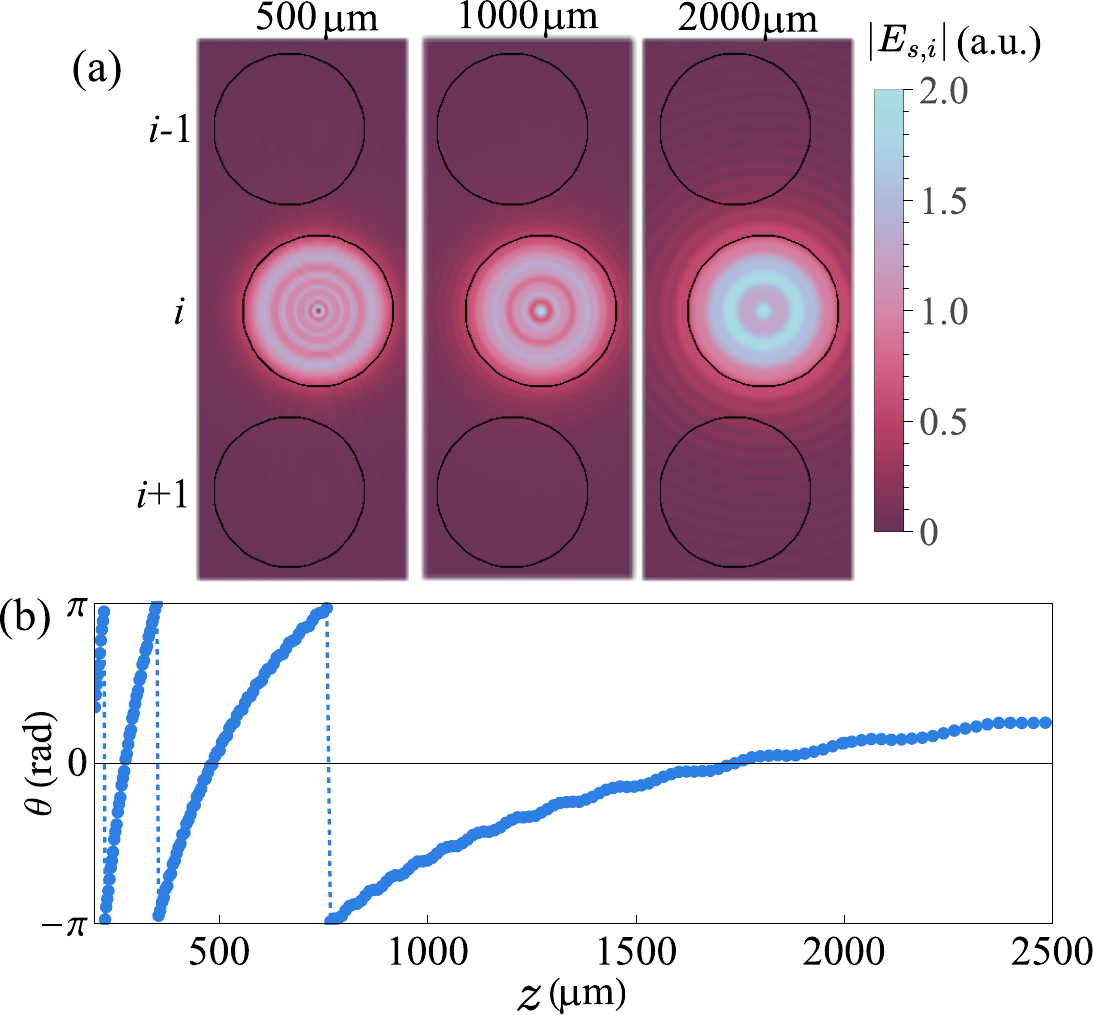}
    \caption{(a) Field amplitude $|E_{s,i}|$ calculated from Eq.\,(\ref{eq:1}) for $z = 500\,\si{\um}, 1000\,\si{\um}$, and $2000\,\si{\um}$ and diameter of the $i$th hole $\sigma_{i} = 200\,\si{\um}$. The distance (center-to-center) between two neighboring is equal to $d=250\,\si{\um}$. (b) The argument of the normalized overlap coefficient as a function of $z$.}
    \label{fig:wf}
\end{figure}

Figure\,\ref{fig:wf}(a) shows the modulus of the field $E_{s,i}$ diffracted from laser number $i$,  calculated with different distances between the mask and OC, namely $z=500\,\si{\um}, 1000\,\si{\um}$, and $2000\,\si{\um}$. The edges of the Uniform mask are marked as black circles. The period of the field oscillations increases with an increase in the distance $z$ between mask and OC. It is easy to understand then that the argument of the complex coupling can be modified by the variation of the mask hole dimensions. 

Figure\,\ref{fig:wf}(b) shows the dependence of the argument $\theta_{i\rightarrow i\pm1}$ of the field coupling coefficient $\kappa_{i\rightarrow i\pm1}$ on the diffraction path length $z$, i.e., the distance between mask and OC. Both holes labeled $i$ and $i\pm 1$ have diameters of $200\,\si{\um}$.
The argument of the overlap varies very fast for small distances $z$. It is thus expected to be hard to control the phase-locking in this range, because of the easy break of symmetry of the laser array through the tilts of the mask, and hole diameter variations. A very precise control of parameters would be needed in this case. The distances $z$ longer than $1\,\si{mm}$ are characterized by much slower variations of the coupling argument, and thus robust phase-locking conditions even with significant asymmetries of the laser array. Due to these reasons, we study the effect of the Gradient and Random masks with a diffraction range $z$ larger than $1\,\si{mm}$.

\begin{figure*}[ht!]
    \centering  
    \includegraphics[width=0.8\linewidth]{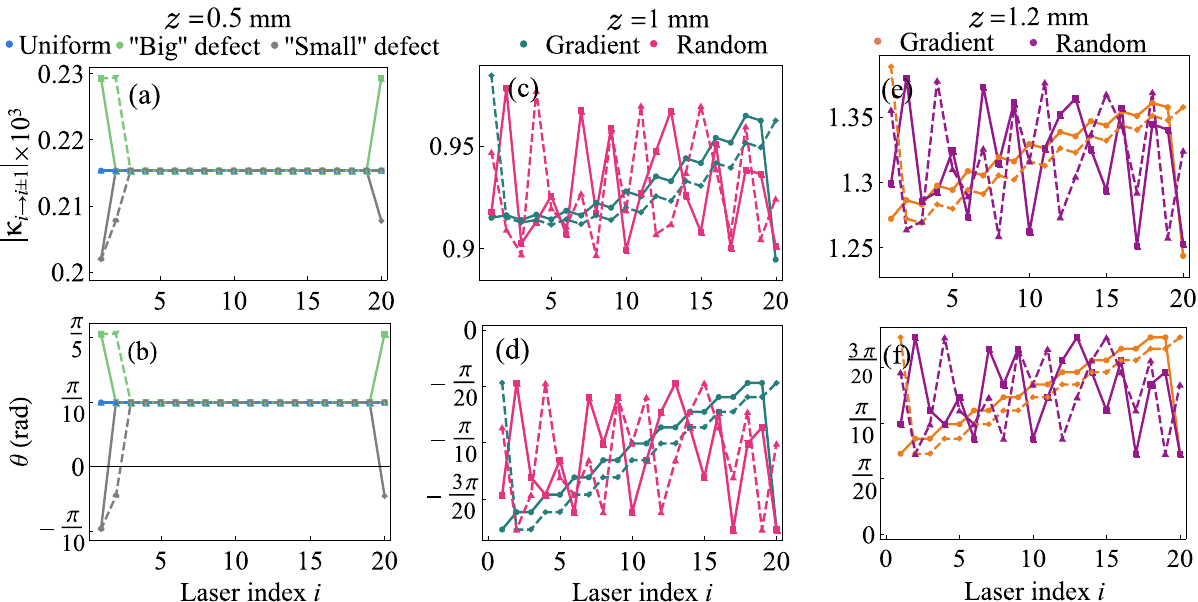} 
      \caption{(a,c,e) Modulus and (b,d,f) argument of the coupling coefficients between neighboring lasers for the different masks.  (a,b) Uniform and Point defect masks studied with $z=500\,\si{\um}$; (c,d) Gradient and Random masks studied both with $z=1000\,\si{\um}$ and  $z=1200\,\si{\um}$ (e,f). Solid lines: $\kappa_{i\rightarrow i+1}$ ; dashed lines: $\kappa_{i\rightarrow i-1}$.}
    \label{fig:eta}
\end{figure*}

We then calculate the coupling coefficient for different coupling geometries and different distances between the mask and output coupler. Figure\,\ref{fig:eta} shows the computed values of the modulus and argument of the coupling coefficients $\kappa_{i\rightarrow i+1}$ (solid lines) and $\kappa_{i\rightarrow i-1}$ (dashed lines) versus laser index $i$ for the different schemes shown in Fig.\,\ref{fig:masks}, with the values of the hole diameters given in Table\,\ref{tab:data} and distance between two neighbouring holes (center-to-center) $d=250\,\si{\um}$. 
%\onecolumngrid\
%\twocolumngrid\

In the case of the Uniform mask (see Figs.\,\ref{fig:eta}\,(a,b)), the coupling coefficient is independent of $i$. The ``Big" and ``Small" defect masks respectively lead to an increase or a decrease of the coupling of the first hole with its neighbors. 
%{\color{red}[I think it should be opposite, as for the same distance z, the light from small hole diffract much faster than bigger one, so injection should be more in case of small hole diameter] [Please check it?]}.
%We elliminate the mask with a uniform field, biger hole will corespond to the bigger total intencity. Smaller hole gives a bigger difraction angle, but the insignificant with a comparision to the increased losses on the mask apperture. 

As expected, the Random mask (see Figs.\,\ref{fig:eta}(c)-\ref{fig:eta}(f)) leads to random variations of the coupling coefficient across the different lasers, while the Gradient mask leads to a progressive variation of the coupling with $i$. Figures\,\ref{fig:eta}(b,d,f) underline how the variations of the hole diameters affect not only the modulus of the coupling but also it's phase. We believe that these variations of $\theta_{i\rightarrow i\pm1}$ will favor the solutions exhibiting a non-zero topological charge, as we investigate in the following sections.

\section{Description of the laser dynamics}
To describe the dynamics of the VECSEL array and analyze its phase-locking behavior, we use a standard rate equation approach. Thanks to the class-A nature of the VECSEL, one can eliminate adiabatically the population inversion dynamics \cite{first}.  
In the system of $n$ lasers, the laser labeled by $i \in [1,n]$ is described by a complex field amplitude $E_i = A_i(t)\; \mathrm{e}^{\textrm{i}\phi_i(t)}$, where $A_i(t)$ and $\phi_i(t)$ are its real-valued amplitude and phase, respectively. In the ring array configuration, each laser field obeys the following set of coupled equations.
\onecolumngrid\
\begin{eqnarray}
   \dv{A_i}{t} &=& -\frac{A_i}{2\tau_{\mathrm{cav}}}\left(1 - \frac{r_{i}}{1 + A_i^2/F_{\mathrm{sat}}}\right)  \nonumber\\
   &+& \frac{c}{2L_{\mathrm{cav}}} \left[ |\kappa_{i+1\rightarrow i}|A_{i+1}\cos(\phi_{i+1} - \phi_i+\theta_{i+1\rightarrow i}) + |\kappa_{i-1\rightarrow i}|A_{i-1}\cos(\phi_{i-1} - \phi_i+\theta_{i-1,i})  \right]\ ,\nonumber\\
    \dv{\phi_i}{t} &=& \Omega_i+\frac{\alpha}{2 \tau_{\mathrm{cav}}} \frac{r_{i}}{1 + A_i^2/F_{\mathrm{sat}}} \nonumber\\
   &+&\frac{c}{2L_{\mathrm{cav}}}\left[|\kappa_{i+1\rightarrow i}|\frac{A_{i+1}}{A_i} \sin(\phi_{i+1} - \phi_i+\theta_{i+1\rightarrow i}) + |\kappa_{i-1\rightarrow i}|\frac{A_{i-1}}{A_i} \sin(\phi_{i-1} - \phi_i+\theta_{i-1\rightarrow i})\right]\
 .\ \label{eq:rate}
\end{eqnarray}
\noindent\rule{\linewidth}{0.4pt}
\vspace{0.2em}
\twocolumngrid\
In these $2n$ coupled differential equations, $\tau_{\mathrm{cav}}$ is the photon lifetime, $r_i$ the excitation ratio of laser $i$, $\Omega_i$ the frequency detuning between lasers, $c$ the speed of light, and $L_{\mathrm{cav}}=4(f_1+f_2)$ the cavity round-trip length. The real amplitude $A_i$ is normalized in such a way that its square corresponds to the photon number in the corresponding laser. Here, $\alpha$ is the Henry factor and $F_{\mathrm{sat}}$ is the saturation photon number, whose exact value serves as a scaling factor for the laser power and does not affect the laser dynamics. We consider the system with zero detuning (same value of $\Omega_i$ for all lasers). In the experimental implementation, this happens in a perfectly degenerate system, thanks to identical cavity lengths and other laser array parameters. This simplification corresponds to a critical coupling value equal to zero \cite{PhysRevA.47.4287}. Thus, the phase-locking can be observed for small values of the coupling coefficients $\kappa_{i\rightarrow \pm1}$.
  
In the numerical simulations described below, we have considered the parameters from our recent experiment \cite{Karuseichyk:23}, with a cavity length $L_{\mathrm{cav}} = 0.5\,\mathrm{m}$ in which the photon lifetime is equal to $\tau_{\mathrm{cav}} = 30\,\mathrm{ns}$.

The steady-state solution of such a system of differential equations is known for exhibiting a multistability of solutions \cite{first}. If the chosen mask is the Uniform one, the steady-state of the system can be easily obtained. Identical coupling parameters lead to phase-locking with the same amplitude all over the array and $A_i=A_{\text{st}}$ for every $i$ given by Eq.\,(\ref{eq:Eq03}) below. The phase-differences are also identical ($\psi_{i}=\psi_{\text{st}}$) and given by Eq.\,(\ref{eq:Eq04}):
\begin{align}
    A_{\text{st}} =&  \sqrt{F_{\mathrm{sat}}}\sqrt{\frac{r}{1-2|\kappa| \displaystyle \frac{c}{L_{\text{cav}}}\cos \psi_{\text{st}}\cos{\theta}}-1}\ ,\label{eq:Eq03}\\
    \psi_{\text{st}}=&  \frac{2 \pi q}{n}\ ,\label{eq:Eq04}
\end{align}
where $q \in\mathbb{Z}$ is often referred to as the topological charge. Phase-locking with $|q|=1$ corresponds to the simplest case of optical vortex, with $2\pi$ phase-shift accumulated around the center of the beam. 

\begin{figure}[h]
    \centering
    \includegraphics[width=1.0\linewidth]{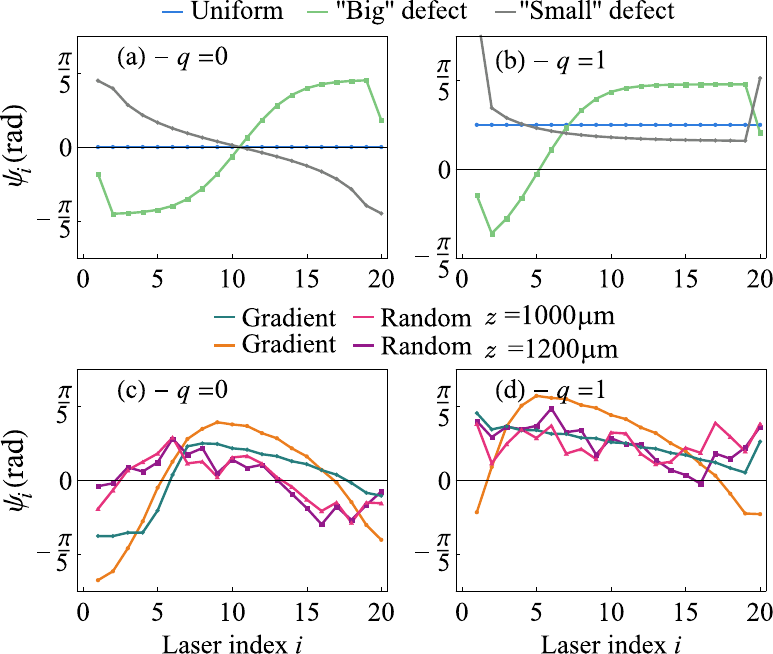}
    \caption{Numerically calculated steady-state phase differences between
    adjacent lasers for the different masks, calculated with $r=1.2, \alpha=2, \tau_{\mathrm{cav}} = 30\,\mathrm{ns}$ and $F_{\mathrm{sat}}=10^{10}$. }
    \label{Fig05}
\end{figure}

Since the system exhibits multistability, the steady-state solution and the associated topological charge that the system will eventually reach strongly depend on the initial conditions. It is possible to obtain a vortex solution for all of the mask configurations within several runs of the rate-equation with different initial conditions. For example, Fig.\,\ref{Fig05} shows the values of the phase differences $\psi_i = \phi_{i+1}-\phi_i$ between successive lasers for two different sets of initial conditions, leading to $q=0$, without vortex (see Fig.\,\ref{Fig05}(a)), and $q=1$, which is a vortex solution (see Fig.\,\ref{Fig05}(b)).  
Indeed, we can see that the numerically calculated $\psi_i$'s are close to zero in Fig.\,\ref{Fig05}(a) and to $2\pi/20$  in Fig.\,\ref{Fig05}(b), as expected from the analytical solutions of Eq.\,(\ref{eq:Eq04}) with $q=0$ and $1$.

It is important to mention that we define the topological charge as $q=\sum_{i=1}^{n}\textrm{arg}(E^*_{i}E_{i+1})/2\pi$ \cite{Pal2015}. For all kinds of non-uniform masks, we can see that the steady-state phase differences no longer satisfy Eq.\,(\ref{eq:Eq04}). It is particularly interesting to have a closer look at what happens with the point defect masks (see Figs.\,\ref{Fig05}\,(a) and \ref{Fig05}\,(b)). We can see that a variation of the diameter of a single hole in the mask results in a redistribution of the perturbation all over the laser array even though only the nearest neighbors are coupled to that particular hole. %{\color{red}[Please add text corresponding to Figs.\,\ref{Fig05}\,(c) and \ref{Fig05}\,(d)]}
Figure\,\ref{Fig05}(c) and (d) show the same redistribution effect for the Gradient and Random masks. One notices that a gradual variation of the hole sizes leads to  smooth variations of the relative phase from hole to hole, while random variations do not.

\begin{figure}[]
    \centering  
    \includegraphics[width=\linewidth]{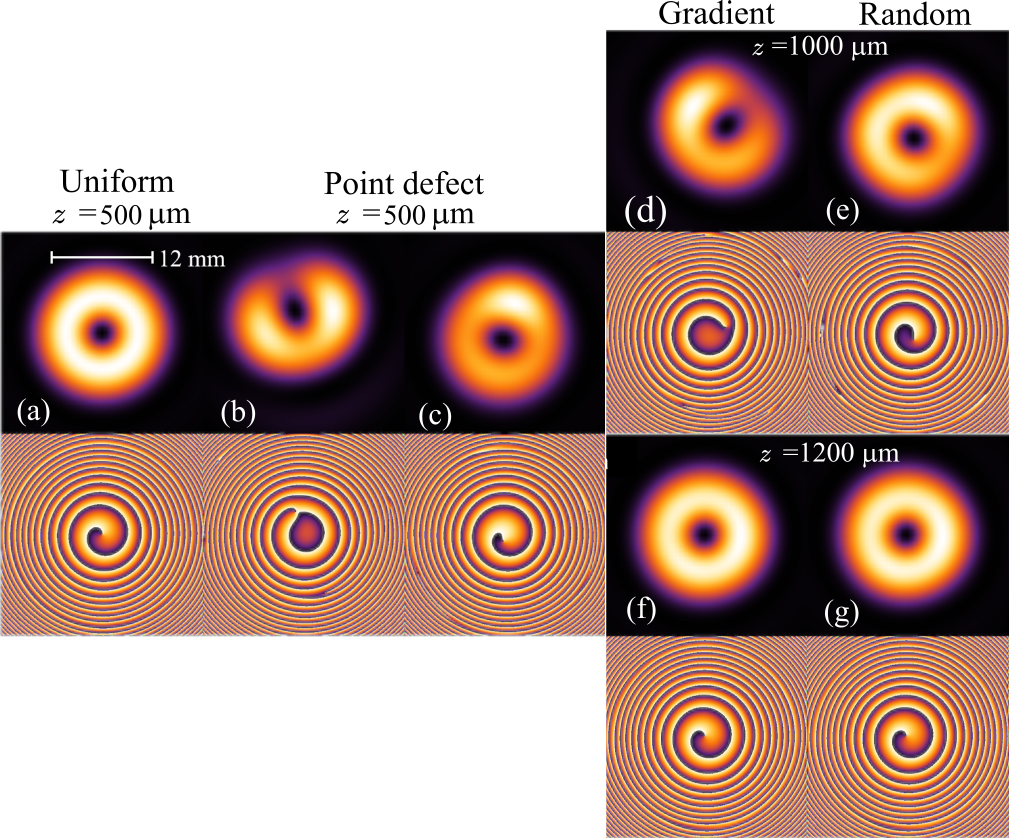}
    \caption{Intensity (top row) and phase (bottom row) patterns of the phase-locked lasers in a ring array of $n=20$ lasers with $q=1$. (a)-(c): results corresponding to ``Uniform", ``Big" defect, and ``Small" defect mask profiles with $z=500\,\si{\um}$. (d) and (f): results for the Gradient mask at $z=1000\,\si{\um}$ (upper) and $z=1200\,\si{\um}$ (lower). (e) and (g): results for the Random mask with $z=1000\,\si{\um}$ (upper) and $z=1200\,\si{\um}$ (lower). The phase maps display the phase variations from 0 to $2\pi$ and the patterns correspond to the solution of the rate equations with coupling parameters shown in Fig.\,\ref{Fig05}. }
    \label{fig:phaseST}
\end{figure}

Using the calculated amplitudes and phase differences of the numerically determined steady-state solutions, we can visualize the far-field patterns of the VECSEL array as shown in Fig.\,\ref{fig:phaseST}.  
The doughnut shape and phase singularity are preserved with the non-uniform masks but with a strongly varying amount of distortion. By comparison with the case of the Uniform mask of Fig.\,\ref{fig:phaseST}(a), it is easy to see that the vortices shown in Figs.\,\ref{fig:phaseST}(c), (f), and (g) have a less distorted phase profile than the other ones. These cases correspond to a mean coupling argument $\theta_{i\rightarrow i\pm1}$ closer to $2\pi/n$ than the other cases. 

The patterns in Figs.\,\ref{fig:phaseST}(b,d,e) show more asymmetric beam structures with complicated phase profiles. In particular,  their singularity point is shifted from the center of the beam because of the strong variations of the coupling argument around its mean value $\theta_{i\rightarrow i\pm1}$. For example, the strongest deviation of the coupling strength is found for the ``Big" defect mask between the lasers labeled $i=20,1$, and 2, where the defect hole corresponds to $i=1$. This leads to a large spatial shift of the topological defect from the center of the beam, as can be seen in Fig.\,\ref{fig:phaseST}(b). 

Each intensity pattern presented in Figs.\,\ref{fig:phaseST}(b)-\ref{fig:phaseST}(g) can be useful as an asymmetric optical vortex. As these AOVs are obtained by phase-locking a ring array of lasers, the output power of the system can be larger than conventional AOVs obtained by a single laser. Thus, these high-power AOVs can be potentially useful in various applications as mentioned earlier. % \st{It has been shown that better results in particle trapping and micro-motoring  can be achieved with an AV beam.} 
 For example, the rate of micro-particle motion is shown to increase linearly with the asymmetry of vortex-carrying Bessel- or Laguerre-Gaussian beams \cite{10.1063/1.4958309, Kovalev:16}. The thermal damage of the live cells is also reported to be better managed when the OV symmetry is broken \cite{10.1063/1.4958309}. Another example of application lies in the fact that vortex-based information storage using the topological charge can be improved by encoding additional information in the non-symmetrical intensity profile \cite{BeamCod}. 
%https://www.researchgate.net/publication/368847289_Tailoring_Large_Asymmetric_Laguerre-Gaussian_Beam_Array_Using_Computer-Generated_Holography
%\st{The shape of the beam is indeed conserved through propagation and thus constitutes a unique parameter, similar to the phase profile of the beam. A strong asymmetry can thus allow  to increase the storage capacity of the beam.}
%bad
%\st{However, a strong deformation of the optical vortex can also make the information carried by the phase profile more difficult to read.}

\section{Limiting value of the Henry factor}
The system has a chance to end up phase-locked in a vortex solution by itself, depending on the initial conditions. However, stable phase-locking cannot be observed for every set of parameters of the system. Among these parameters, a peculiarity of semiconductor lasers is the presence of a relatively large Henry factor. In preceding works, it was established that this factor limits the probability to obtain a vortex solution \cite{first}. In the case of weak real positive coupling, it was shown that the probability for the laser array to exhibit a vortex solution with $|q|=1$ becomes negligible if the Henry factor increases above limiting value $\alpha_{\text{lim}}=2/\text{tan}\,\psi_q$.
However, this investigation was limited to real-valued coupling coefficients, and thus needs to be re-considered for the complex coupling coefficients that we consider here.

We thus follow the same approach to obtain a more general expression for the complex coupling case. First of all, we linearise the rate equations given by Eqs.\,(\ref{eq:rate}) around the steady-state solutions of Eqs.\,(\ref{eq:Eq03}) and (\ref{eq:Eq04}). In this way, we obtain the Jacobian of the system, from which the stability of the steady-state solution can be analyzed based on its trace and its determinant. A given steady-state solution is stable if the trace is negative and the determinant positive.

The details of the calculations are given in the Appendix. They show that the condition on the trace is always fulfilled for our system. Furthermore, the condition on the determinant leads to the following stability criterion: stable phase-locking with $\psi_q$ is possible only when the $\alpha$ factor of the gain chip is smaller than a  limit value given by:
\begin{equation}
 \alpha_{\text{lim}}=\frac{2 \cos \theta\cos\psi_q}{ \sin\left( \psi_q-\theta\right)}.
    \label{eq:RLcond4}
\end{equation}
The limitations are different for different topological charges and vary with the number of lasers. Additionally, we can see a clear dependency of the limiting Henry factor on the coupling argument. 

\begin{figure}[htbp]
    \centering
    \includegraphics[width=\linewidth]{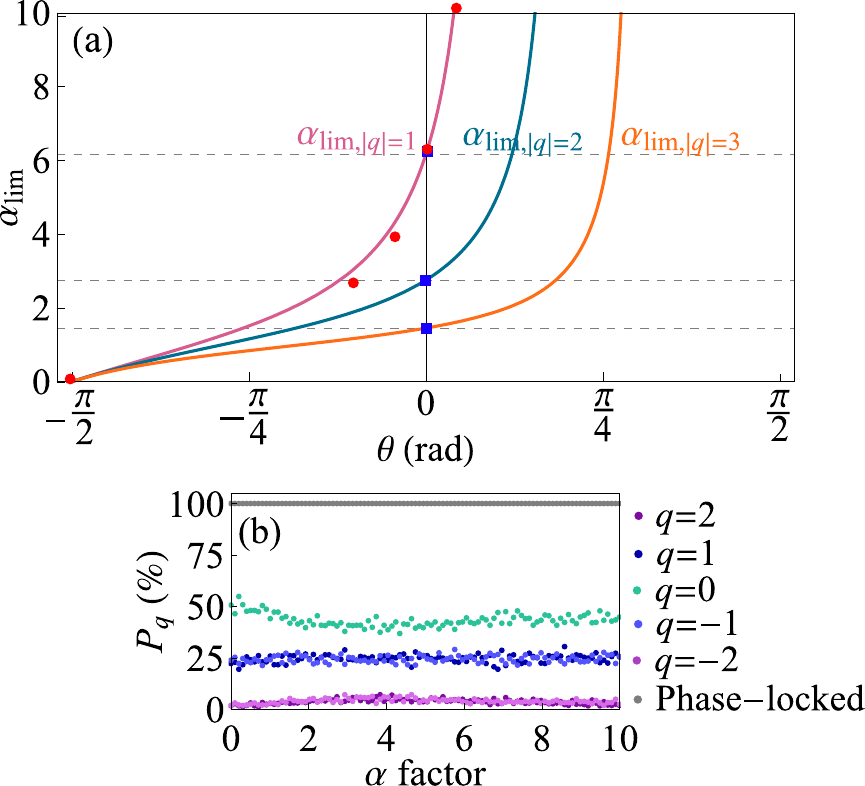}
    \caption{(a) Evolution of the limiting values of the Henry factor versus argument of the coupling coefficient for $|q|=1$ (pink line), $|q|=2$ (blue line), $|q|=3$ (orange line), according to Eq.\,(\ref{eq:RLcond4}). Horizontal dashed lines correspond to a real positive coupling ($\theta=0$). The blue squares are the values predicted by the formula in Ref. \cite{first}. Red circles: values obtained from numerical simulations for $|q|=1$ and $\theta=-\pi/2,-\pi/10,-\pi/20$, and $\pi/20$. (b) Probability to phase-lock the laser array with the value of the topological charge $q$ for the Uniform mask at $z=500\,\si{\um}$, as a function of $\alpha$. The coupling argument is taken to be $\theta=\pi/10$.} 
    \label{fig:ALassumtions}
\end{figure}

A graphical illustration of Eq.\,(\ref{eq:RLcond4}) is given in Fig.\,\ref{fig:ALassumtions}(a) for $|q|= 1, 2$ and $3$. The horizontal dashed lines correspond to the limiting values of the Henry factor predicted in Ref. \cite{first} for the real coupling factor, i.e. $\theta=0$, as given by the blue squares.    
We then compare the analytical formula of Eq.\,(\ref{eq:RLcond4}) with numerical simulations in the case of a Uniform mask with different values of the coupling argument ($\theta=-\pi/2, -\pi/10,-\pi/20,\pi/20,\pi/10$ and $\pi/2$) and with $|\kappa_{i\rightarrow i\pm 1}|=0.28\times10^{-3}$. We calculate the number of outcomes phase-locked with $|q|=1$ among the 500 runs of the rate equations of Eqs.\,(\ref{eq:rate}) with different initial conditions. From this, we obtain the value of $\alpha$ above which the probability to obtain a steady-state solution with $|q|=1$ becomes negligible.  These simulations lead to the red dots in Fig.\,\ref{fig:ALassumtions}(a), which perfectly confirm the analytical result of Eq.\,(\ref{eq:RLcond4}). 

Figure \ref{fig:eta}(a) has shown that the Uniform mask leads to a coupling argument  $\theta=\pi/10$ when $z=500\,\si{\um}$. We thus take this value to compute the probabilities of the different solutions shown in Fig.\,\ref{fig:ALassumtions}(b). As expected from Fig.\,\ref{fig:ALassumtions}(a), no limit value for the Henry factor was observed in this case. Indeed, the analytically predicted value for this case exceeds 10. This shows that if one is able to control the value of the argument of the coupling coefficient, then steady-state solutions carrying a non-zero topological charge can be obtained with significant probability even with large values of the Henry factor.

\section{Phase-locking probability with non-uniform loss-mask profile}
\begin{figure}[htbp]
\centering
    \includegraphics[width=\linewidth]{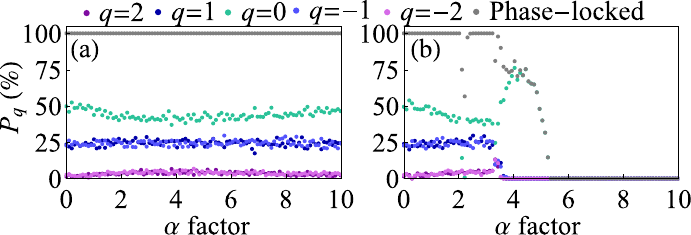}
    \caption{Probabilities for the laser array to exhibit phase-locking with different topological charges $q$ as a function of $\alpha$, for (a) ``Big" and (b) ``Small" defect masks. Coupling coefficients are calculated for $z=500\,\si{\um}$, as shown in Figs.\,\ref{fig:eta}(a,b).}
    \label{fig:prob}            
\end{figure}

In this section, we investigate whether the probabilities to generate phase-locked vortex solutions with $|q|=1$ and 2 can be increased by using different non-uniform masks. The probabilities obtained for the ``Big" and ``Small" Point defect masks with the coupling coefficients presented in  Figs.\,\ref{fig:eta}(a,b) are shown in Fig.\,\ref{fig:prob}. 
They are calculated with $500$ random initial conditions for each value of $\alpha$. The probabilities for the Uniform (see Fig.\,\ref{fig:ALassumtions}(b)) and ``Big" defect (see Fig.\,\ref{fig:prob}(a)) masks are very close. In both cases, the argument of the coupling coefficient has a positive sign for each pair of lasers in the array. The limit value for the $\alpha$ factor is very large in both cases. 
         
A very different result is obtained for the ``Small'' defect mask, as can be seen in Fig.\,\ref{fig:prob}\,(b).  In this case, the limit value of the Henry factor is approximately $\alpha_{\text{lim}}=3.8$, which is very small compared to the preceding cases. Only the $q=0$ solution appears to be stable in the range  $3.8\leq\alpha\leq5.3$. There is an interesting effect close to $\alpha=2$, where 100\% of the phase-locked cases lead to a non-zero topological charge. The width of this region, in which the $q=0$ experiences large losses, increases with a decrease of the coupling strength $|\eta|$, but the physical explanation of this effect is not fully clear.
%{\color{red}[It seems that the inphase solution becomes very lossy and does not lase in this parameter regime. We can add some text based on these.]}.
%
None of the masks investigated in Fig.\,\ref{fig:prob} leads to an asymmetry between solutions corresponding to positive and negative values of $q$ and the probabilities of finding OV with positive and negative TCs remain equal.

\begin{figure}[htbp]
    \centering
    \includegraphics[width=\linewidth]{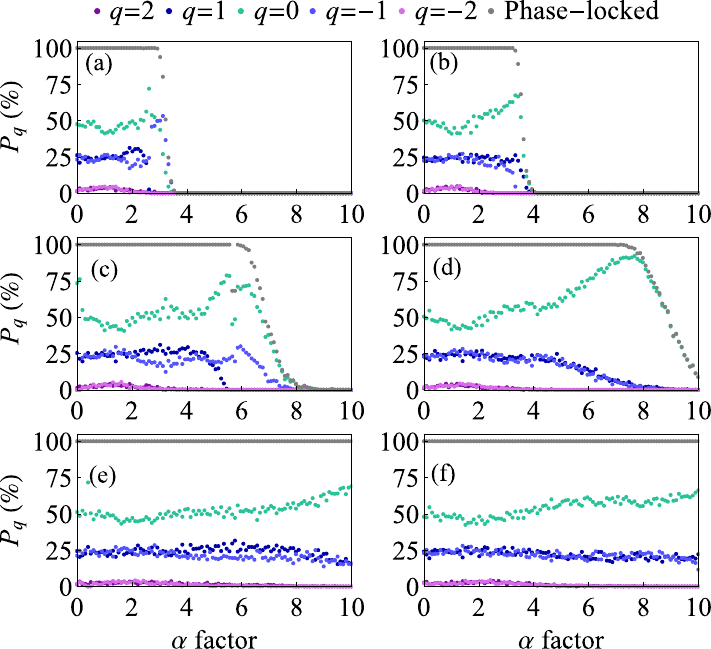}
    \caption{Probabilities for the laser array to exhibit phase-locking with different topological charges $q$ as a function of $\alpha$, for (a,c,e) Gradient and (b,d,f) Random masks. The results are obtained for (a,b) $z=1050\,\si{\um}$, (c,d) $z=1100\,\si{\um}$, and (e,f) $z=1200\,\si{\um}$.}
    \label{fig:GRprob}
\end{figure}

The behavior of the laser array with the Gradient and Random masks, calculated for three different values of the distance $z$, is shown in Fig.\,\ref{fig:GRprob}.
The first two rows of the figure are calculated for $z=1050\,\si{\um}$ and $1100\,\si{\um}$, a situation in which the signs of $\theta_{i\rightarrow i\pm1}$ can be either positive or negative depending on the considered pair of lasers. This makes the average of these arguments over the array to be equal approximately to $-\pi/60$ and $\pi/25$ in cases (a) and (b), respectively. Thus the phase-locking ranges decrease or increase with respect to the preceding case, as expected from Eq.\,(\ref{eq:RLcond4}). The arguments of the coupling coefficients are all positive in the cases of Figs.\,\ref{fig:GRprob}(e,f), which have been calculated for $z=1200\,\si{\um}$. %
%In[1491]:= Mean[Flatten@{\[Theta]Gm3, \[Theta]Gp3}]
%Out[1491]= -0.13953
%In[1492]:= Mean[Flatten@{\[Theta]Gm4, \[Theta]Gp4}]
%Out[1492]= 0.0570578
%In[1493]:= Mean[Flatten@{\[Theta]Rm3, \[Theta]Rp3}]
%Out[1493]= -0.13953
%In[1494]:= Mean[Flatten@{\[Theta]Rm4, \[Theta]Rp4}]

Interestingly, the results for the Gradient mask exhibit a break of symmetry (unequal probability) between positive and negative values of $q$ for $2.5\leq \alpha \leq 2.8$ in Fig.\,\ref{fig:GRprob}(a) and  $5.8$ to $\leq \alpha \leq 8$ in Fig.\,\ref{fig:GRprob}(c). This is a promising result for the generation of anti-vortices. Also, a small range of values of $\alpha$ around $0.4$ exhibits destabilization of the $q=1$ solution, leading consequently to a symmetry breaking between positive and negative vortices in Fig.\,\ref{fig:GRprob}(c). This breaking of the symmetry just gives an advantage of a few percent for the probability $P_{q=1}$ over $P_{q=-1}$. 

The effect of the random mask is less spectacular. Let us first focus on the comparison of the data presented in Figs.\ref{fig:GRprob}(a) and \ref{fig:GRprob}(b), which respectively correspond to the Gradient and Random masks. Interestingly, the last one exhibits a narrow range of values of $\alpha$ ($3.8\leq\alpha\leq 4$), in which $P_{q=1}>P_{q=-1}$. The dominant solution is still $q=0$, but the degeneracy between positive and negative topological charges is lifted. At the same time, the probability of phase locking of the laser array decreases dramatically in this range.

The distribution of the hole diameters in the Random mask averages the non-uniformity of the mask parameters, thus explaining the fact that the symmetry-breaking range is significantly smaller in this case. 

Comparing Figs.\,\ref{fig:GRprob}(e) and \ref{fig:GRprob}(f) show that only the Gradient mask leads to a predominance of the vortex solution over the anti-vortex one, however with a probability difference of only a few percents.  Overall, the most promising mask geometry is the Gradient one. The best positioning strategy is to first consider the limiting $\alpha$ factor effect within the $\alpha$ range of the used gain chip. The average of $\theta_{i\pm1\rightarrow i}$'s can then be obtained using Eq.\,(\ref{eq:RLcond4}), and the distance $z$ can be determined from Fig.\,\ref{fig:wf}(b) or from the calculation based on the specific mask being used.

\begin{figure}[htbp]
    \centering
    \includegraphics[width=\linewidth]{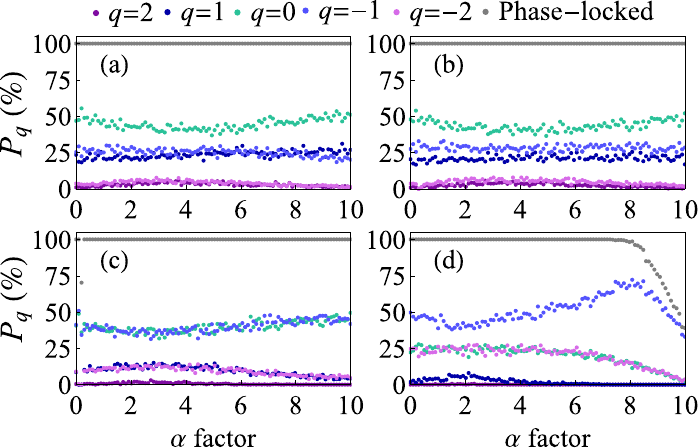}
    \caption{Probabilities to obtain topological charge with a specially designed mask satisfying the criteria $\theta_{i\rightarrow i-1}=2\pi/n$ and $\theta_{i\rightarrow i+1}=-2\pi/n$ for (a)  holes $i=20,1,2$, (b) holes $i=20,1,2,3$, (c)  half of the array and (d) in which each hole satisfies the condition. Zero holes satisfying these criteria correspond to the probability shown in Fig.\,\ref{fig:prob}\,(a).}
    \label{fig:probAs}
\end{figure}

In the next step, we consider the case of a virtual mask, that adds phase shifts $\theta_{i \rightarrow i+1}=2\pi/n$ and $\theta_{i\rightarrow i-1}=-2\pi/n$ between lasers to force the system to phase lock according to the anti-vortex solution $q=-1$. The result is shown in Fig.\,\ref{fig:probAs}. Figure \ref{fig:probAs}(a) corresponds to the case where $\theta_{i \rightarrow i+1}=2\pi/n$ and $\theta_{i\rightarrow i-1}=-2\pi/n$ is valid only for $i=1$, while for the other lasers, this argument keeps the value obtained for the Uniform mask. In Fig.\,\ref{fig:probAs}(b) this condition is valid for $i=1$ and $i=2$. It is extended to half of the lasers, i.e. for $1\leq i\leq 10$ (see Fig.\,\ref{fig:probAs}(c)) and to the entire array (see Fig.\,\ref{fig:probAs}(d)).

The results of Fig.\,\ref{fig:probAs} show that such masks would indeed allow to increase the value of $P_{q=-1}$, and to lift the degeneracy between positive and negative values of the topological charge. Interestingly, this increase of $P_{q=-1}$ occurs thanks to a decrease of $P_{q=0}$. Moreover, when one increases the number of lasers in the array for which $\theta_{i \rightarrow i\pm1}=\pm 2\pi/n$, the probability $P_{q=-1}$ increases, as it can be seen by comparing the successive plots in Fig.\,\ref{fig:probAs}. In particular, when the number of lasers for which $\theta_{i \rightarrow i+1}$ is equal to $\pm 2\pi/n$ exceeds $n/2$, one obtains $P_{q=-1}\ge P_{q=0}$. In this case, the solution with non-zero topological charge becomes dominant, as can be seen for example in Fig.\,\ref{fig:probAs}(d).

Finally, let us mention that the generation of asymmetric OVs with higher values of $q$ would probably be possible by increasing the number of lasers $n$.%{\color{red}[We have not said anything on generating asymmetric OVs with higher values of $q$. We can add text saying that by increasing the number of lasers n, one can also obtain OV with higher TC.]}
%\textcolor{green}{We also studied the hight-values ($q=9, 8, \cdot$) AV in the case, which becomes dominant steady-states with a negative coupling argument. Such steady-states are characterized by a very complicated unstructured far-field profile. Such solutions we found inapplicable for the above mentioned applications.}

\section{Conclusions}
We present a new approach for asymmetric vortex beam generation by phase-locking ring array of lasers in a DVECSEL. Particularly, we explore the impact of non-uniform masks on the phase-locking of a ring array of lasers, taking into account the complex nature of the coupling coefficients and the existence of a non-zero Henry factor. We have shown that the phase-locking of  laser array, and the predominance of the solution in which all the lasers are in phase, are relatively robust to variations of the hole diameters with respect to their average values. However, we have also observed that strong deviations of the hole diameters are detrimental to the phase-locked operation of the laser array. For example, in the case of an amplitude mask containing holes with random diameters (Random mask), the standard deviation in hole diameters of average diameter equal to $200\,\mu\mathrm{m}$ should not exceed approximately $2\,\si{\um}$ at a distance $z=1$\,mm and $5\,\si{\um}$ for $z=2$\,mm. This also makes the fabrication of the mask by standard machining techniques a bit challenging.

%\st{Contrary to our initial expectations, We have observed that the non-uniform masks that we have considered, and in particular the initially promising Gradient mask profile, do not demonstrate significant benefits for vortex beam generation with respect to standard uniform masks.} 
We have observed that the in-phase solution always dominates irrespective of non-uniform masks. Moreover, we have found that the quality of the vortex beam, and its asymmetries, are strongly dependent on the complex coupling argument.

Further, we have predicted a possibility to make one particular vortex solution dominant over other possible steady-state solutions. This solution consists in imprinting the necessary phase shift among neighboring lasers in the argument of their coupling coefficients, namely $\theta_{i,i+1}=2\pi/n$ and $\theta_{i+1,i}=-2\pi/n$. In particular, a larger number of holes satisfying this condition is favorable for the establishment of a given vortex solution. However, the design of a mask allowing to fulfill this condition for a significant number of holes in the array remains to be found. 

We believe that this method can prove useful to select one desired vortex phase-locked solution in the laser array, in conjunction if necessary with far-field Fourier filtering. However, this can be possible only through further studies of ``diffraction engineering", to some extent related to what is required for laser neuron network training and control \cite{MiriMenon+2023+883+892}, which we plan to explore in the future work.

This work was partially supported by the PAUSE program and by QuanTEdu-France. VP acknowledge the funding from Science and Engineering Research Board (SIR/2022/00019 and CRG/2021/003060).

\section*{DATA AVAILABILITY}
The data that support the findings of this study are available from the corresponding author upon reasonable request.
%
%\bibliography{ref1}% Produces the bibliography via BibTeX.

\section*{APPENDIX. RING LASER ARRAY STABILITY ANALYSIS}
This appendix provides additional details on the derivation of the analytical formula given by Eq.\,(\ref{eq:RLcond4}). This formula characterizes the maximal Henry factor $\alpha_{\textrm{lim}}$ of the laser gain chip allowing the ring laser array to be phase-locked in a vortex. This is a result of the stability analysis of the phase-locking ring laser arrays. 

A clear picture of the system's stability is given by the analysis of its Jacobian. The rank of the Jacobian increases as $2n-1$, where $n$ is the number of lasers. Then, the analysis requires a study of the $2n-1$ eigenvalues, which is not easy to compute even for $n=20$ lasers.

However, for the simplification, we can isolate the sub-system in one of the lasers interacting with its neighbors \cite{first}. Let us consider the steady-state amplitude and phase of the selected laser $\{A_q, \psi_q\}$ with a variation of its parameters around this steady-state solution according to the rate-equations Eq.\,(\ref{eq:rate}). 

The Jacobian matrix obtained by linearization of Eq.\,(\ref{eq:rate}) around steady-state is then:

\begin{align}
&\mathbf{J}(A_q,\psi_{q}) \equiv \left.
\begin{bmatrix}
 \displaystyle \frac{\partial}{\partial A_i} \left ( \displaystyle \dv{A_i}{t} \right )  & \displaystyle \frac{\partial}{\partial \psi_i} \left ( \dv{A_i}{t} \right ) 
 \vspace{0.5em}\\
 \displaystyle \frac{\partial}{\partial A_i} \left ( \dv{\psi_i}{t} \right ) & \displaystyle \frac{\partial}{\partial \psi_i} \left ( \dv{\psi_i}{t} \right ) 
\end{bmatrix}
\right\rvert_{A_q, \psi_q}
\end{align}

The steady-state solution is stable if all eigenvalues of $\mathbf{J}(A_i,\psi_{i})$ have a negative real part. When applied to the presented Jacobian, this is equivalent to claim that $\mathrm{Tr}(\mathbf{J}) <0 $ and $\mathrm{Det}(\mathbf{J}) >0$. 
Here we consider these criteria separately.

\subsection{Trace of the Jacobian}
The condition for the negative Jacobian trace has the form of Eq.\,(\ref{eq:condDet}), when the calculated trace is simplified as
\begin{align}
\mathrm{Tr}(\mathbf{J} )=\frac{1}{\tau_{\textrm{cav}}}\frac{4|\eta|^2\cos^2\theta\cos^2\psi_{q}-4\eta\cos\theta\cos\psi_{q}}{r} \nonumber \\  +\frac{1}{\tau_{\textrm{cav}}}\frac{1-r}{r}<0, \, \label{eq:condDet}
\end{align} 
This condition is a quadratic equation of $|\eta|$ and leads to the following requirement for the phase-locking:
\begin{align}
-\frac{1}{2} \left(\sqrt{r}-1\right)\frac{1}{\cos\psi _q} < |\eta|\cos\theta <  \frac{1}{2} \left(\sqrt{r}+1\right)\frac{1}{\cos\psi _q}. \label{eq:RLcond1}
\end{align}
If the coupling is real ($\theta=0$), this condition is always true \cite{first}. Additionally, it is worth mentioning that the factor $\sqrt{r}\pm1$ is always positive when the pumping rate is above the laser threshold ($r>1$). The argument of the coupling varies between $-\pi$ and $\pi$, same as the phase difference between lasers $\psi_q$. 
Since the condition defined by the trace is always true the stability can thus be limited only by the determinant of the Jacobian matrix. 

\subsection{Determinant of the Jacobian}
Let us now consider the condition determined by the Jacobian determinant. Its formula can be simplified considering a small coupling strength, which is typical for a small length $z$ of the diffraction cavity:
\begin{align}
 \mathrm{Det}(\mathbf{J})=\frac{1}{\tau_{\textrm{cav}}}\frac{|\eta|  (r-1)}{2r}\left [  \cos \psi _q (\alpha  \sin \theta+2 \cos\theta)\right.\nonumber\\+\left.\alpha  \cos\theta \sin \psi _q\right ]+\mathcal{O}\left(\eta ^2\right) > 0.  
\label{eq:condDet3}
\end{align} 
From this, we obtain a new requirement, which says that the stable phase-locking with $\psi_q$ is possible with $\alpha<\alpha_{\text{lim}}$ given by:
\begin{equation}
\alpha_{\text{lim}}=\frac{2 \cos \theta\cos\psi_q}{ \sin\left( \psi_q-\theta\right)}.
    \label{eq:RLcond4_2}
\end{equation}
It is clear that the limitations are different for different topological charges ($\psi_q=2\pi q/n$). The limiting value of the $\alpha$ factor is $\pi$ periodic in $\theta$. A graphical illustration of the formula is given in Fig.\,\ref{fig:ALassumtions}\,(a) for $|q|= 1, 2, 3$ where the coupling argument $\theta$ is in a range from $-\pi/2$ to $\pi/2$. 

We can see that the choice of $\theta$ crucially affects the stability range for each presented TC ($q$). The $\alpha$ range below the $q$-curves corresponds to the stability of this phase-locked solution. Conversely, the $\alpha$ factor values above these curves correspond to a negligible probability of observing the corresponding topological charge. 
Equation\,(\ref{eq:RLcond4_2}) can be simplified for a long diffraction cavity length $z$, where the coupling argument $\theta$ tends to 0. Then, the limiting Henry factor is given by \cite{first}:
\begin{align}
    \alpha_{\text{lim}}=\frac{ 2}{\displaystyle\tan{\frac{2\pi q}{n}}},
    \label{eq:alphaAlit}
\end{align}
and its simplified expression given by a Taylor expansion is:
   \begin{align}
   \alpha_{\text{lim}} =\frac{2}{\tan{\displaystyle \frac{2 \pi}{n}}} = \frac{n}{\pi} - \frac{2}{3} \frac{2 \pi}{n} + \mathcal{O} \left(\frac{1}{n^3}\right).
   \label{eq:alphLim}
\end{align}

\end{document}